\documentclass[trackchanges, twocolumn]{aastex701}
\usepackage{comment}
\usepackage{amsmath} 
\usepackage{bm}
\usepackage{mathrsfs}
\usepackage{subcaption}
\usepackage{afterpage}
\usepackage{placeins}

\begin{document}

\title{Prompt Jet Launching and Delayed Thermal Disk in Tidal Disruption Event for Prolonged Activity GRB~220706A}

\author[orcid=0000-0003-2477-9146]{Yuri Sato}
\affiliation{Astronomical Institute, Graduate School of Science, Tohoku University, Sendai 980-8578, Japan}
\email[show]{yuri@astr.tohoku.ac.jp}

\author[orcid=0009-0000-1257-5133]{Rin Oikawa}
\affiliation{Astronomical Institute, Graduate School of Science, Tohoku University, Sendai 980-8578, Japan}
\email[show]{oikawa.rin@astr.tohoku.ac.jp}

\begin{abstract}

GRB~220706A exhibits X-ray flaring activity extending to $\sim51$~days after the trigger, along with a luminous optical thermal component emerging on a timescale of $\sim10^6$~s. 
Such prolonged central-engine activity could be powered by a collapsar fed by an extended stellar envelope.
However, a relativistic jet may struggle to break out of the extended envelope.
These challenges motivate a tidal disruption event (TDE), although explaining both the early relativistic jet and the luminous late optical thermal emission within one TDE framework remains difficult.
We investigate a TDE scenario in which shocks between stellar debris streams drive the early inflow of some debris, potentially carrying preexisting fossil magnetic flux toward the black hole (BH) and powering an initial relativistic jet via the Blandford–Znajek mechanism.
The remaining debris circularizes into an accretion disk, whose thermal emission accounts for the late optical thermal component.
For GRB~220706A, the optical timescale, luminosity, and color temperature are consistent with disruption of a blue supergiant ($M_*\sim50~M_\odot$, $R_*\sim20~R_\odot$) by a BH of mass $M_{\rm BH}\sim10^5~M_\odot$, assuming that the accretion disk forms on a timescale comparable to the orbital period of the most tightly bound stellar debris.
A magnetic field strength of $\sim10^5$~G is required near the BH before disruption to power a jet capable of producing the inferred gamma-ray luminosity. 
Continued fallback may sustain central-engine activity beyond the observed X-ray flaring episode, making future X-ray monitoring a test of this scenario.

\end{abstract}


\section{Introduction}\label{Introduction}

Gamma-ray bursts (GRBs) are relativistic transients commonly associated with the launch of relativistic jets from accreting compact objects \citep[e.g.][]{Piran2004,Zhang2019}. 
GRB~220706A has a prompt gamma-ray duration of $\sim87$~s, typical of the long GRB population. 
On the other hand, follow-up X-ray observations revealed multiple flaring episodes extending to approximately 51 days after the trigger \citep{Gompertz2026}, indicating exceptionally prolonged central-engine activity. 
Late time observations also revealed luminous optical thermal emission \citep{Gompertz2026}. 
The origin of the prolonged X-ray activity and the physical mechanism responsible for the late time optical thermal emission remain unclear.

One of the leading models for the central engines of long GRBs is the collapsar scenario, in which the core collapse of a massive star forms a compact object whose accretion powers a relativistic jet \citep[e.g.][]{Woosley1993,MacFadyen1999,Woosley2006}.
In this framework, the duration of central-engine activity is governed by the timescale over which stellar material continues to accrete onto the central object.
To explain the exceptionally prolonged X-ray activity observed in GRB~220706A, a collapsar involving a progenitor with an extended stellar envelope may be a plausible candidate.
However, for such a progenitor, a relativistic jet may struggle to break out of the stellar envelope before the central engine shuts off \citep[e.g.][]{Bromberg2011,Suwa2011,Nakauchi2013,Bromberg2015}.
Furthermore, energy deposited by the jet into the cocoon may heat and expand the stellar envelope, potentially suppressing further accretion onto the central compact object \citep{Quataert2012}.

An alternative approach to addressing these difficulties is to consider a scenario in which a relativistic jet is launched from a black hole (BH) without an extended stellar envelope.
A promising candidate is a tidal disruption event (TDE), in which a star is disrupted by the tidal forces of a BH and the returning stellar debris supplies material for accretion onto the BH \citep{Hills1975,Rees1988,EK1989}.
The fallback of stellar debris can provide a prolonged supply of accreting material to power the central engine, consistent with the extended activity of GRB~220706A.
However, explaining both the launching of an early relativistic jet and the luminous optical thermal emission emerging on a timescale of $\sim10^6$~s within a single TDE scenario remains challenging \citep{Gompertz2026}.

To address this issue, we investigate a TDE scenario involving early magnetic flux transport followed by the formation of an accretion disk. 
In this scenario, shocks between stellar debris streams induce an early inflow of a fraction of the debris, which transports pre-existing magnetic flux toward the BH \citep{Tripto2026}. 
This magnetic flux can enable the launching of an initial relativistic jet through the Blandford--Znajek (BZ) mechanism \citep{Blandford1977}. 
Meanwhile, the subsequently returning debris circularizes to form the main accretion disk, whose thermal emission may account for the late optical transient.

In this Letter, we first assess the viability of the collapsar scenario for GRB~220706A by examining whether a relativistic jet can break out of the progenitor’s extended stellar envelope within the available central-engine lifetime (Section~\ref{sec:Core-collapse}). 
We then investigate the TDE scenario (Section~\ref{sec:model}), constraining the properties of the BH-star system using the characteristic timescale, luminosity, and color temperature of the late optical thermal emission. 
The magnetic field strength required near the BH horizon to power the relativistic jet via the BZ mechanism, as well as the corresponding field strength in the preexisting disk, is also estimated.
Finally, the duration of jet activity sustained by fallback accretion is evaluated, and the implications for future X-ray observations are discussed.

\section{Core collapse scenario} 
\label{sec:Core-collapse}

Observations suggest that the central engine of GRB~220706A remains active for more than 10 days \citep{Gompertz2026}. 
To investigate whether such prolonged activity can be explained by a core-collapse scenario, we first consider a core-collapse scenario in which the central engine is powered by accretion of the stellar envelope. Assuming that the envelope falls toward the center on the free-fall timescale, the characteristic timescale can be estimated as
\begin{equation}
\begin{aligned}
t_{\rm ff}
&= \frac{1}{\sqrt{G\bar{\rho}}}
= \sqrt{\frac{4\pi R_*^3}{3GM_*}} \\
&\sim  10^6\: \left(\frac{R_*}{200\:R_{\odot}}\right)^{\frac{3}{2}}\left(\frac{M_*}{15\:M_{\odot}}\right)^{-\frac{1}{2}}\: {\rm s}\;,
\label{eq:free_fall}
\end{aligned}
\end{equation}
where $M_*$, $R_*$, and $\bar{\rho}$ are the stellar mass, radius, and mean density, respectively. Here $G$ is the gravitational constant. 
For a red giant with $M_*\sim15\:M_\odot$ and $R_*\sim200\,R_\odot$, the free-fall timescale can be sufficiently long to motivate this type of progenitor as a candidate for sustaining prolonged central-engine activity. 
However, this interpretation requires not only a sufficiently long accretion timescale but also successful jet breakout through the extended stellar envelope. We examine the latter requirement below.

\begin{figure*}[t]
\centering
\includegraphics[width=\textwidth]{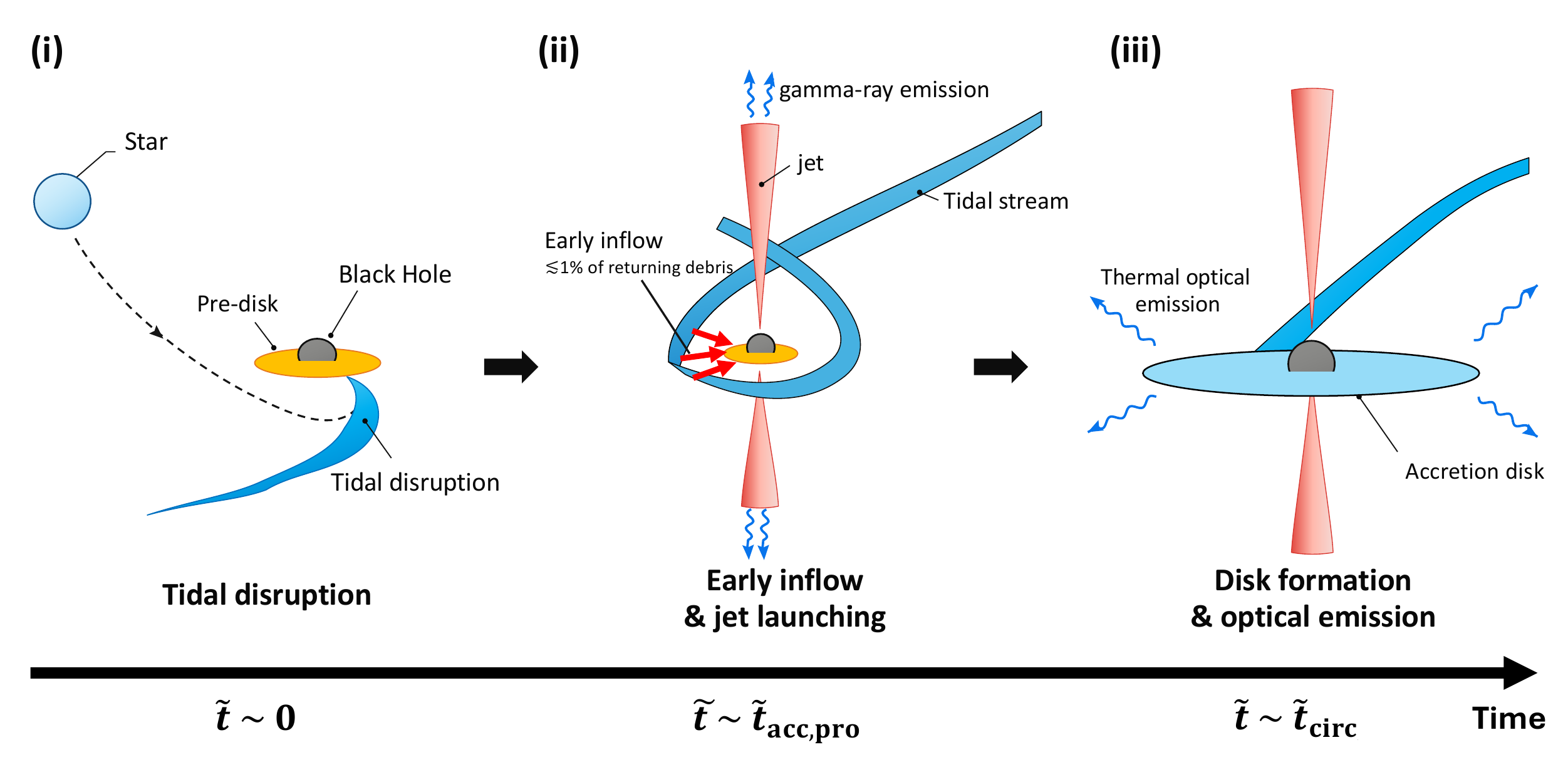}
\caption{Schematic illustration of the proposed TDE scenario for GRB~220706A. 
(i) A star is tidally disrupted by a BH surrounded by a preexisting magnetized disk ($\tilde{t}=0$). 
(ii) As the stellar debris returns to pericenter, a small fraction of the debris ($\lesssim1\%$) flows inward, transporting magnetic flux from the preexisting disk toward the BH and potentially powering an early relativistic jet via the BZ mechanism at $\tilde{t}\sim\tilde{t}_{\rm acc,pro}$.
(iii) At later times, the returning debris circularizes to form the main accretion disk, whose thermal emission accounts for the delayed optical thermal component observed on a timescale of $\tilde{t}_{\rm opt,obs}\sim10^6~{\rm s}$.}
\label{fig:theory_picture}
\end{figure*}

Following core collapse, a BH surrounded by an accretion disk is expected to form at the center and power a GRB jet.
Two representative mechanisms have been proposed for launching GRB jets: a fireball powered by neutrino–antineutrino annihilation above the accretion disk \citep{Eichler1989,Meszaros1993,Zalamea2011}, and the BZ mechanism, which electromagnetically extracts the rotational energy of the BH \citep{Blandford1977}.
Although a BZ jet may initially be magnetically dominated, magnetic-energy dissipation inside the stellar envelope may allow it to approach a thermally dominated hydrodynamic state (see Appendix~\ref{appendix} for details). We therefore adopt a hydrodynamical jet to estimate the jet breakout time.
If the central engine ceases its activity before the jet head reaches the stellar surface, the jet fails to break out of the stellar envelope \citep{Bromberg2011,Suwa2011,Nakauchi2013,Bromberg2015}. 
The jet-head velocity can be estimated from the balance of momentum flux between the jet and the surrounding stellar material \citep{Bromberg2011}:
\begin{equation}
\beta_{\rm h}
= \frac{\beta_{\rm j}}{1+\tilde{L}^{-\frac{1}{2}}},
\label{eq:jet_head_velocity}
\end{equation}
where $\beta_{\rm j}$ is the jet velocity normalized by the speed of light $c$, and
\begin{equation}
\tilde{L}
= \frac{L_{\rm j}}{S\rho_*c^3}.
\label{eq:dimensionless_luminosity}
\end{equation}
Here, $L_{\rm j}$ is the jet luminosity, $S=\pi R_{\rm j}^2\sim\pi\theta_j^2r^2$ is the jet cross-sectional area, $\rho_*$ is the density of the stellar material, $\theta_j$ is the jet half opening angle, $R_j$ is the cylindrical radius of the jet, measured perpendicular to its axis, and $r$ is the radial distance from the BH. 
The breakout time required for the jet head to reach the stellar surface is then estimated as
\begin{equation}
t_{\rm bo}
= \int_{R_{\rm c}}^{R_*}\frac{dr}{\beta_{\rm h}c},
\label{eq:breakout_time}
\end{equation}
where $R_{\rm c}$ is the stellar core radius. If the central engine is powered primarily by an accretion disk formed from the stellar core, its activity timescale is approximately determined by the viscous timescale of the disk:
\begin{equation}
\begin{aligned}
t_{\rm engine}
&= \frac{1}{\alpha_{\rm vis}}
\sqrt{\frac{R_{\rm c}^3}{GM_{\rm BH}}} \\
&\sim5\times10^2\,\left(\frac{\alpha_{\rm vis}}{0.1}\right)^{-1}\left(\frac{R_{\rm c}}{10^{10}\:{\rm cm}}\right)^{3/2}\\
&\:\:\:\:\:\times\left(\frac{M_{\rm BH}}{3\:M_{\odot}}\right)^{-1/2}\,\rm s,
\end{aligned}
\label{eq:engine_time}
\end{equation}
where $\alpha_{\rm vis}$ is the viscosity parameter. For a red giant, we adopt an envelope density profile of $\rho_*\propto r^{-2.5}$, an envelope mass of $M_{\rm env}\approx11\,M_\odot$, and a core mass and radius of $M_{\rm c}\approx4\,M_\odot$ and $R_{\rm c}\approx10^{10}\ {\rm cm}$, respectively \citep{Suwa2011}. We also adopt the jet power inferred for GRB~220706A,
\begin{equation}
\begin{aligned}
L_{\rm j}
&= \frac{\theta_j^2}{2}L_{\rm j,iso}\\
&\sim 5\times10^{47}
\left(\frac{L_{\rm j,iso}}{10^{50}\ {\rm erg\ s^{-1}}}\right)
\left(\frac{\theta_j}{0.1\ {\rm rad}}\right)^2~{\rm erg\ s^{-1}},
\end{aligned}
\end{equation}
where $L_{\rm j,iso}$ is the isotropic-equivalent jet luminosity. 
Using these parameters, we find $t_{\rm bo}\sim 10^4\:{\rm s}$ and
\begin{equation}
t_{\rm engine}<t_{\rm bo}.
\label{eq:engine_breakout}
\end{equation}
In this case, the central engine shuts off before the jet head reaches the stellar surface, preventing the jet from breaking out of the extended envelope.

In addition, the propagation of the jet through the stellar envelope can affect the subsequent mass supply to the central engine. 
While the jet is confined within the star, a substantial fraction of its energy is deposited into the cocoon through the jet head \citep{Bromberg2011}. 
The accumulated cocoon energy can heat and expand the stellar envelope and may eventually unbind a significant fraction of it \citep{Quataert2012}. 
Consequently, the envelope material that would otherwise accrete onto the central engine on the free-fall timescale estimated in Equation~\eqref{eq:free_fall} may instead be expelled, potentially suppressing the mass supply required to sustain prolonged activity.
Taken together, the difficulty of achieving jet breakout within the available engine lifetime and the potential suppression of accretion by cocoon feedback pose challenges for the collapsar scenario as an explanation for GRB~220706A.

\section{Tidal Disruption Scenario} 
\label{sec:model}

In the previous section, we showed that it is difficult for a relativistic jet to break out of the extended stellar envelope required to explain the prolonged central-engine activity of GRB~220706A in the core-collapse scenario. 
An alternative possibility is a TDE, in which a star is disrupted by the tidal force of a BH. 
Unlike in the collapsar scenario, a jet in a TDE scenario need not propagate through an extended stellar envelope surrounding the central engine, potentially alleviating the jet breakout problem. 
In addition, the fallback of stellar debris can sustain accretion onto the BH over an extended period, potentially explaining the prolonged central-engine activity. 
A TDE interpretation has also been proposed for the recently discovered ultra-long GRB~250702B \citep{Levan2025,Beniamini2025,Eyles-Ferris2025,EP2025,Granot2025,Yuan2026,Sato2026}.
Then, we explore whether GRB~220706A can be powered by the tidal disruption of a star of mass $M_*$ and radius $R_*$ by a BH of mass $M_{\rm BH}$.

A key challenge for this TDE scenario is to explain both the early launching of a relativistic jet and the delayed emergence of thermal optical emission.
We propose a scenario in which this time delay is explained by the early inflow of stellar debris and the subsequent formation of the main accretion disk. Figure~\ref{fig:theory_picture} illustrates a schematic picture of this scenario. The evolution of the TDE can be divided into three main stages:

(i) The star is tidally disrupted during its first pericenter passage by a BH surrounded by a preexisting magnetized disk.

(ii) As the disrupted stellar debris returns to pericenter, a small fraction flows inward toward the BH. This inflow transports magnetic flux from the pre-disk toward the BH, enabling the launching of a relativistic jet via the BZ mechanism. 

(iii) At later times ($\sim10^6\,{\rm s}$), the stellar debris circularizes to form the main accretion disk, producing thermal optical emission. 

We investigate whether the early jet activity and the delayed thermal optical emission can be explained within a TDE framework, and constrain the properties of the BH-star system using the characteristic timescale, luminosity, and color temperature of the optical emission, as well as the observed gamma-ray luminosity.

\subsection{Tidal Disruption of the Star}
\label{sec:label}

We consider a star moving on a Keplerian orbit around a BH. The star is tidally disrupted when its pericenter distance $R_p$ becomes smaller than the tidal radius $R_T$, given by
\begin{equation}
\begin{aligned}
R_T
&=R_*\left(\frac{M_{\rm BH}}{M_*}\right)^{\frac{1}{3}}\\
&\sim 10^{13}
\left(\frac{M_{\rm BH}}{10^5 \:M_\odot}\right)^{\frac{1}{3}}
\left(\frac{M_*}{50\:M_\odot}\right)^{-\frac{1}{3}}\\
&\:\:\:\:\:\:\times
\left(\frac{R_*}{20\:R_\odot}\right)
\,{\rm cm}.
\end{aligned}
\end{equation}
Following tidal disruption, the stellar debris is divided into gravitationally bound and unbound components. The bound debris follows eccentric orbits and eventually returns to the vicinity of the BH \citep{Rees1988,EK1989}. We define the penetration factor as $\beta=R_T/R_p\gtrsim1$. Assuming that the energy spread of the debris is determined at pericenter, the fallback timescale of the most tightly bound debris is given by \citet{Rees1988} as
\begin{equation}
\begin{aligned}
t_{\rm fb}
&=\frac{\pi}{\sqrt{2}}\beta^{-3}
\left(\frac{GM_*}{R_*^3}\right)^{-\frac{1}{2}}
\left(\frac{M_{\rm BH}}{M_*}\right)^{\frac{1}{2}} \\
&\sim 10^6\,\beta^{-3}
\left(\frac{M_{\rm BH}}{10^5\:M_\odot}\right)^{\frac{1}{2}}
\left(\frac{M_*}{50\:M_\odot}\right)^{-1}
\left(\frac{R_*}{20\:R_\odot}\right)^{\frac{3}{2}}
\,{\rm s}.
\end{aligned}
\end{equation}
Hereafter, we introduce a new time coordinate $\tilde{t}$, with its origin set at $t=t_{\rm fb}$:
\begin{equation}
\tilde{t}=t-t_{\rm fb}.
\end{equation}

\subsection{Jet Production in the Early Phase}
\label{sec:jet}

The most tightly bound stellar debris returns to the vicinity of the BH after the fallback timescale $t_{\rm fb}$. Recent numerical simulations suggest that angular-momentum redistribution at the nozzle shock deflects only a small fraction of the returning debris inward, allowing only $\lesssim 1\%$ of the debris to be promptly accreted by the BH \citep{Ryu2023,Steinberg2024,Price2024,Krolik2025,Abolmasov2026}. 
We estimate the characteristic timescale for this prompt accretion as
\begin{equation}
\begin{aligned}
\tilde{t}_{\rm acc,pro}
&=\frac{1}{\alpha_{\rm vis}}
\sqrt{\frac{R_p^3}{GM_{\rm BH}}}\\
&\sim 10^5\,\beta^{-\frac{3}{2}}
\left(\frac{\alpha_{\rm vis}}{0.1}\right)^{-1}\\
&\:\:\:\:\times
\left(\frac{M_*}{50\:M_\odot}\right)^{-\frac{1}{2}}
\left(\frac{R_*}{20\:R_\odot}\right)^{\frac{3}{2}}\:{\rm s}.
\end{aligned}
\label{eq:prompt_accretion_timescale}
\end{equation} 
This timescale characterizes the early inflow required for jet launching and will be compared with the circularization timescale in Section~\ref{sec:disk} to assess whether the jet can be launched before the main accretion disk forms.
If a preexisting magnetized disk is present around the BH, a small fraction of the stellar debris flowing inward may interact with the pre-disk and transport its magnetic flux toward the horizon, potentially enabling the launching of a relativistic jet through the BZ mechanism \citep{Tripto2026}.

We estimate the magnetic field strength $B_H$ at the BH horizon resulting from the magnetic flux transported through this process. We assume a radial magnetic-field profile in the pre-disk as
\begin{equation}
B(R)=B_0\left(\frac{R}{r_H}\right)^{-q},
\end{equation}
where $B_0$ is the magnetic field strength at $r_H$ extrapolated from the pre-disk profile, and $q$ characterizes the radial dependence of the magnetic field. The magnetic flux contained between the BH horizon and the pericenter distance $R_p$ is then estimated as
\begin{equation}
\begin{aligned}
\Phi_{\rm disk}
&\simeq \int_{r_H}^{R_p}2\pi R B(R)\,dR\\
&=\frac{2\pi B_0r_H^2}{2-q}
\left[
\left(\frac{R_p}{r_H}\right)^{2-q}-1
\right].
\end{aligned}
\end{equation}
Assuming that this magnetic flux is efficiently transported to and accumulated on the BH horizon, such that
\begin{equation}
\Phi_{\rm disk}\approx\Phi_H=2\pi r_H^2B_H,
\end{equation}
we obtain the magnetic field strength at the horizon as
\begin{equation}
B_H\simeq
\frac{B_0}{2-q}
\left[
\left(\frac{R_p}{r_H}\right)^{2-q}-1
\right].
\label{eq:Bfield}
\end{equation}
If sufficient magnetic flux is supplied to the BH horizon, the early relativistic jet can be explained by the BZ mechanism, even before the formation of the main accretion disk from the tidally disrupted stellar debris \citep{Tripto2026}. 

The isotropic-equivalent gamma-ray luminosity produced by the BZ jet is given by \citet{Oikawa2026}
\begin{eqnarray}
\label{eq:L_X}
L_{\rm \gamma,iso,obs}&\sim&\frac{4c\eta_{\rm BZ}\eta_{\rm rad,\gamma}r_H^2B_H^2}{\theta_j^2}\nonumber\\
&\sim&10^{50}~x_H^2\left(\frac{\eta_{\rm BZ}}{0.1}\right)\left(\frac{\eta_{\rm rad,\gamma}}{0.1}\right)\left(\frac{\theta_j}{0.1~{\rm rad}}\right)^{-2}\nonumber\\
&&\times\left(\frac{M_{\rm BH}}{10^5~M_\odot}\right)^{2}\left(\frac{B_H}{10^{8}~{\rm G}}\right)^{2}~{\rm erg~s^{-1}},
\end{eqnarray}
where $r_H=\left(1+\sqrt{1-a_{\rm spin}^2}\right)r_g=x_Hr_g$ is the horizon radius, $r_g=GM_{\rm BH}/c^2$ is the gravitational radius, $a_{\rm spin}$ is the dimensionless BH spin parameter, $\eta_{\rm BZ}$ is the effective efficiency of the BZ mechanism, $\eta_{\rm rad,\gamma}$ is the radiative efficiency for converting the jet power into gamma-ray emission, and $\theta_j$ is the jet opening angle.

\subsection{Disk Formation in the Late Phase}
\label{sec:disk}

In the previous subsection, we showed that the early inflow of stellar debris returning to pericenter may transport magnetic flux from the preexisting magnetized disk to the BH horizon, potentially powering a relativistic jet during the early stage of the TDE. Here, we discuss the late optical thermal emission in the TDE scenario.

In our scenario, the late optical thermal emission originates from the newly formed main accretion disk.
The physical processes governing the circularization of tidally disrupted stellar debris and their characteristic timescales remain poorly understood.
We therefore parameterize the characteristic circularization timescale, measured from the time at which the most tightly bound debris returns to pericenter as
\begin{equation}
\tilde{t}_{\rm circ}=f_{\rm dis}t_{\rm fb},
\end{equation}
where $f_{\rm dis}$ is a dimensionless parameter. We assume that the stellar debris circularizes efficiently and adopt $f_{\rm dis}=1$.
The condition for the jet to be launched at an early stage, followed by the formation of an accretion disk at a later stage, is that the initial accretion timescale is shorter than the circularization timescale $\tilde{t}_{\rm acc,pro}<\tilde{t}_{\rm circ}$.
This condition imposes the following constraint on the BH mass $M_{\rm BH}$ and stellar mass $M_*$:
\begin{equation}
\left(\frac{M_*}{50~M_\odot}\right)
<
25
\left(\frac{\alpha_{\rm vis}}{0.1}\right)^2
\left(\frac{M_{\rm BH}}{10^5~M_\odot}\right)
f_{\rm dis}^2\beta^{-3}.
\label{eq:condition}
\end{equation}
Within the parameter space satisfying this inequality, the jet can be launched through the initial accretion of stellar debris before circularization, while the subsequent formation of an accretion disk can account for the late optical thermal emission.

The characteristic timescale of the optical thermal emission is determined by the circularization timescale of the stellar debris,
\begin{eqnarray}
\label{eq:t_tot}
\tilde{t}_{\rm opt,obs} &\sim& (1+z)\tilde{t}_{\rm circ}\sim(1+z)f_{\rm dis}t_{\rm fb} \nonumber\\
&\sim& 2\times10^6\left(\frac{1+z}{1.8577}\right)\left(\frac{M_{\rm BH}}{10^5~M_\odot}\right)^{\frac{1}{2}}\left(\frac{M_*}{50~M_\odot}\right)^{-1}\nonumber\\
&&\times\left(\frac{R_*}{20~R_\odot}\right)^{\frac{3}{2}}f_{\rm dis}\beta^{-3}~{\rm s},
\end{eqnarray}
where $z$ is the source redshift. Assuming that the debris conserves its specific angular momentum during circularization, the circularization radius is estimated as
\begin{equation}
R_{\rm c}\approx2R_p=\frac{2R_{\rm T}}{\beta}.
\end{equation}

Assuming that the gravitational energy released during the formation of the main accretion disk powers the optical thermal emission, its luminosity can be estimated as
\begin{eqnarray}
\label{eq:L_opt}
L_{\rm opt,obs}&\sim& \eta_{\rm rad,opt}\frac{GM_{\rm BH}\dot{M}_{\rm pk}}{R_c}\nonumber\\
&\sim&10^{44}~\left(\frac{\eta_{\rm rad,opt}}{0.3}\right)\left(\frac{M_{\rm BH}}{10^5~M_\odot}\right)^{\frac{1}{6}}\left(\frac{M_*}{50~M_\odot}\right)^{\frac{7}{3}}\nonumber\\
&&\times\left(\frac{R_*}{20~R_\odot}\right)^{-\frac{5}{2}}~{\rm erg~s^{-1}},
\end{eqnarray}
where $\eta_{\rm rad,opt}$ is the effective radiative efficiency for converting the released gravitational energy into optical emission. Here, $\dot M_{\rm pk}={M_*}/({3t_{\rm fb}})$ is the characteristic peak fallback rate \citep{Tchekhovskoy2014}.

Finally, we estimate the color temperature of the optical thermal emission. Assuming blackbody emission and that the effective photospheric radius is comparable to the circularization radius $R_{\rm c}$, the observed color temperature is estimated as
\begin{eqnarray}
\label{eq:e_opt}
T_{\rm opt,obs}&\sim& \left(\frac{L_{\rm opt,obs}}{4\pi \sigma_{\rm SB} (1+z)^4 R_{\rm c}^2}\right)^{\frac{1}{4}}\nonumber\\
&\sim& 10^4~\left(\frac{1+z}{1.8577}\right)^{-1}\left(\frac{L_{\rm opt,obs}}{10^{44}~{\rm erg~s^{-1}}}\right)^{\frac{1}{4}}\nonumber\\
&&\times\left(\frac{M_{\rm BH}}{10^5~M_\odot}\right)^{-\frac{1}{6}}\left(\frac{M_*}{50~M_\odot}\right)^{\frac{1}{6}}\left(\frac{R_*}{20~R_\odot}\right)^{-\frac{1}{2}}~{\rm K},\nonumber\\
\end{eqnarray}
where $\sigma_{\rm SB}$ is the Stefan–Boltzmann constant.

\subsection{Constraints on the System Parameters}
\label{sec:parameter}

We constrain the BH mass $M_{\rm BH}$ and the stellar mass $M_*$ and radius $R_*$ using the characteristic timescale, extinction-corrected luminosity, and color temperature of the optical component. 
In the following estimates, we adopt $z=0.8577$ and a fiducial optical radiative efficiency of $\eta_{\rm rad,opt}=0.3$, and an extinction-corrected optical luminosity $L_{\rm opt,obs}\sim10^{44}~{\rm erg~s^{-1}}$, estimated using $A_V=3.6$~mag \citep{Gompertz2026}.
Combining Equations~(\ref{eq:t_tot}), (\ref{eq:L_opt}), and (\ref{eq:e_opt}), we obtain the following approximate relations:
\begin{eqnarray}
M_{\rm BH} &\sim& 10^5~M_\odot \left(\frac{\tilde{t}_{\rm opt,obs}}{10^6~{\rm s}}\right)^{3}\left(\frac{T_{\rm opt,obs}}{10^4~{\rm K}}\right)^{4}\nonumber\\
&&\times\left(\frac{\eta_{\rm rad,opt}}{0.3}\right)^{3}f_{\rm dis}^{-\frac{1}{3}}\beta^9,\\
M_*&\sim&50~M_\odot \left(\frac{\tilde{t}_{\rm opt,obs}}{10^6~{\rm s}}\right)^{-2}\left(\frac{T_{\rm opt,obs}}{10^4~{\rm K}}\right)^{-6}\nonumber\\
&&\times\left(\frac{L_{\rm opt,obs}}{10^{44}~{\rm erg~s^{-1}}}\right)^{\frac{3}{2}}\left(\frac{\eta_{\rm rad,opt}}{0.3}\right)^{-6}f_{\rm dis}^{-2}\beta^{-6},\nonumber\\\\
R_*&\sim&20~R_\odot\left(\frac{\tilde{t}_{\rm opt,obs}}{10^6~{\rm s}}\right)^{-\frac{5}{3}}\left(\frac{T_{\rm opt,obs}}{10^4~{\rm K}}\right)^{-\frac{16}{3}}\nonumber\\
&&\times\left(\frac{L_{\rm opt,obs}}{10^{44}~{\rm erg~s^{-1}}}\right)\left(\frac{\eta_{\rm rad,opt}}{0.3}\right)^{-5}f_{\rm dis}^{-\frac{17}{9}}\beta^{-5}.\nonumber\\
\end{eqnarray}
These estimates suggest the TDE involving a massive blue supergiant and an IMBH. 
For this progenitor, the condition in Eq.~\eqref{eq:condition} is satisfied, allowing the jet to be launched during the initial accretion of stellar debris before circularization, followed by the formation of the main accretion disk responsible for the late optical thermal emission.

We next estimate the magnetic field strength near the BH horizon using the observed isotropic-equivalent gamma-ray luminosity. From Equation~\eqref{eq:L_X}, we obtain
\begin{equation}
\begin{aligned}
B_H \sim {}&
10^{8}
\left(\frac{L_{\rm \gamma,iso,obs}}{10^{50}~{\rm erg~s^{-1}}}\right)^{1/2}
\left(\frac{\tilde{t}_{\rm opt,obs}}{10^6~{\rm s}}\right)^{-3}\left(\frac{T_{\rm opt,obs}}{10^4~{\rm K}}\right)^{-4}\\
&\times
\left(\frac{\eta_{\rm BZ}}{0.1}\right)^{-1/2}\left(\frac{\eta_{\rm rad,\gamma}}{0.1}\right)^{-1/2}
\left(\frac{\theta_j}{0.1~{\rm rad}}\right)
x_H^{-1}~{\rm G}.\\
\end{aligned}
\end{equation}
Here, we have eliminated the BH mass $M_{\rm BH}$ using the constraints derived from the optical emission. We then estimate the magnetic field strength required in the pre disk to produce such a strong horizon-scale magnetic field through the magnetic flux stored in the pre-disk. Adopting $q\sim5/4$ \citep{Yuan2014} for the radial magnetic-field profile and $R_c\sim2R_T/\beta$ for the characteristic radius of the pre-disk, Equation~\eqref{eq:Bfield} indicates that a pre-disk magnetic field strength of $B_0\sim10^5~{\rm G}$ is required. If such magnetic flux was already stored in the pre-disk prior to the TDE, a relativistic jet could have been launched even before the tidal disruption. However, Equation~\eqref{eq:L_X} gives the corresponding pre-jet power as
$L_{\rm pre,jet}\sim10^{42}~{\rm erg~s^{-1}}$,
which is significantly lower than the luminosity of the prompt emission from GRB~220706A. At the redshift of the source ($z=0.8577$), any associated X-ray emission may have been too faint to be detected by current X-ray instruments.

\subsection{Prediction for the Jet Shutoff Time}
\label{sec:shut_off}

The jet activity is expected to persist while the accretion flow is super-Eddington and to cease when the accretion rate declines to approximately the Eddington rate, potentially accompanying a transition to a geometrically thin disk \citep{Tchekhovskoy2014,Shen2014}. 
Assuming a fallback accretion rate of
$\dot{M}(\tilde{t})\approx\dot{M}_{\rm pk}(1+{\tilde{t}}/{t_{\rm fb}})^{-5/3}$,
we estimate the jet shutoff time, $\tilde{t}_{\rm off}$, by setting $\dot{M}(\tilde{t}_{\rm off})\approx\dot{M}_{\rm Edd}$, where $\dot{M}_{\rm Edd}$ is the Eddington accretion rate. 
For the fiducial parameters inferred for GRB~220706A, our estimate gives 
\begin{eqnarray}
\tilde{t}_{\rm off}&\sim&10^9~\left(\frac{M_{\rm BH}}{10^{5}~M_\odot}\right)^{-\frac{2}{5}}\left(\frac{M_{*}}{50~M_\odot}\right)^{\frac{1}{5}}\left(\frac{R_{*}}{20~R_\odot}\right)^{\frac{3}{5}}\beta^{-\frac{6}{5}}~{\rm s}.\nonumber\\
\end{eqnarray}
This suggests that the central engine could remain active well beyond the observed X-ray flaring episode. 
Continued X-ray monitoring can test this prediction and further constrain the long-term evolution of the central engine.

\section{Summary \& Discussion}
\label{sec:discussion}

We have investigated the origin of the prolonged high-energy activity and delayed optical thermal emission of GRB~220706A. 
Our results suggest that the collapsar scenario faces difficulty in explaining the observed relativistic jet, which is unlikely to break out of an extended stellar envelope before the central engine shuts off. 
As an alternative, we have proposed the TDE scenario in which an early inflow of stellar debris may power the relativistic jet, while the subsequent fallback and circularization of the debris produce the delayed optical thermal emission. 
By interpreting the characteristic timescale, extinction-corrected luminosity, and color temperature of the optical component in this scenario, we infer characteristic parameters of $M_{\rm BH}\sim10^5\,M_\odot$, $M_*\sim50\,M_\odot$, and $R_*\sim20\,R_\odot$,
{assuming that the accretion disk forms over a timescale comparable to the orbital period of the most tightly bound stellar debris.} 
These results suggest that the TDE of a blue supergiant by an IMBH provides a promising framework for explaining the diverse emission properties of GRB~220706A.

An important uncertainty in our scenario is the circularization timescale of the stellar debris. 
In TDEs, stream self-intersections driven by relativistic apsidal precession have been discussed as a mechanism for accretion disk formation \citep[e.g.,][]{Hayasaki2013,Bonnerot2016}. 
However, {a spinning BH} can cause debris geodesics to deviate from a fixed orbital plane, potentially leading to misalignment between successive debris streams. 
As a result, stream self-intersections may be suppressed, delaying accretion disk formation \citep{Guillochon2015}. 
This uncertainty complicates the interpretation of the late optical thermal emission as thermal radiation from an accretion disk.
In our scenario, a magnetized accretion flow is assumed to be present around the BH before the TDE. Previous MHD simulations have shown that interactions between stellar debris and a preexisting magnetized disk can lead to turbulent mixing and magnetic flux capture \citep{Kelley2014}. 
Such interactions may also facilitate energy dissipation and angular momentum transport through magnetic stresses and shocks, potentially promoting debris circularization even when stream self-intersections are inefficient \citep{Kelley2014,Bonnerot2017}.
{Nevertheless,} the efficiency of this process depends on the magnetic-field geometry, magnetization, and ambient gas density. 
{Moreover,} magnetic fields may inhibit stream mixing and delay energy dissipation under certain conditions \citep{Pacuraru2026}. 
Therefore, although we assume the formation of an accretion disk when modeling the late optical thermal emission, determining its formation timescale requires global MHD simulations that account for interactions between stellar debris and the preexisting magnetized accretion flow in Kerr spacetime.

GRB~220706A exhibits multiple X-ray flares extending to $\sim51$ days after the trigger. 
X-ray flares in GRBs have been proposed to arise from prolonged central engine activity \citep{Lazzati2007,Maxham2009}, although their physical origin remains uncertain. 
In our TDE scenario, the fallback of stellar debris can sustain accretion onto the BH over an extended period, potentially allowing the central engine to remain active and thereby providing a possible context for the observed X-ray flares.

\begin{acknowledgements}

We thank 
Kazumi~Kashiyama,
Takanori~Sakamoto,
Shuta~J.~Tanaka,
Kenji~Toma, 
and
Seiji~Toshikage
for valuable comments.
This research was partially supported by JSPS KAKENHI Grant No.~25KJ0010 (Y.S.).
R.O. was supported by the Graduate Program on Physics for the Universe (GP-PU).

\end{acknowledgements}

\begin{contribution}
The authors contributed equally to this work.
\end{contribution}

\appendix

\section{The Propagation of a Poynting dominated Jet inside the Star}
\label{appendix}

A BZ jet is initially magnetically dominated at the jet base.
As the jet propagates through the stellar envelope, the pressure of the surrounding material collimates the jet and may promote the development of current-driven instabilities (CDIs), which may facilitate magnetic-energy dissipation \citep{Levinson2013,Bromberg2016,Globus2016,Tchekhovskoy2016}. This dissipation may drive the magnetically dominated jet toward a thermally dominated state.
Following the analytical treatment of \citet{Oikawa2026}, we estimate the characteristic distance at which a CDI develops. 
We parameterize the jet geometry as
\begin{equation}
R_j\propto r^\alpha,
\end{equation}
where $\alpha$ characterizes the jet collimation, $R_j$ is the cylindrical radius of the jet, measured perpendicular to its axis, and $r$ is the radial distance from the BH. 
The characteristic distance for CDI development can then be expressed as
\begin{equation}
\frac{r_{\rm CDI}}{r_g}
\simeq
\left(\kappa\alpha\Gamma_j\right)^{\frac{1}{1-\alpha}},
\end{equation}
where $\Gamma_j$ is the jet Lorentz factor, and $\kappa$ is a parameter that depends on the magnetic-field geometry. For a representative parabolic jet with $\alpha=1/2$, adopting $\Gamma_j=10$ and $\kappa=10$ \citep{Oikawa2026}, we obtain
\begin{equation}
r_{\rm CDI}\sim \text{a few}\times10^3\,r_g
\sim10^9~{\rm cm}<R_\odot,
\end{equation}
where we adopt $M_{\rm BH}=3\,M_\odot$. 
Thus, in a star with an extended envelope, such as a red giant, CDIs may develop well before the jet reaches the stellar surface, potentially facilitating magnetic-energy dissipation and allowing the jet to approach a thermally dominated hydrodynamic state.

\bibliography{sample701}{}
\bibliographystyle{aasjournalv7}

\end{document}